\documentclass[oldversion]{aa}
\usepackage{natbib}
\usepackage{graphicx}
\usepackage{txfonts}

\newcommand\refereebf[1]{{\bf #1}}
\renewcommand\refereebf[1]{{#1}}
\newcommand\beq{\begin{equation}} 
\newcommand\eeq{\end{equation}}
\newcommand{\id}{\ensuremath{\,\mathrm d}}
\newcommand{\diff}[3]{\ensuremath{\displaystyle\frac{\id^{#2}#1}{\id {#3}^{#2}}}}

\titlerunning{Blue stragglers in N-body models of M67}
\title{Evolution of stellar collision products in open clusters. I. 
Blue stragglers in N-body models of M67}
\author{
   Evert Glebbeek\inst{1} \and
   Onno R. Pols\inst{1} \and
   Jarrod R. Hurley\inst{2}
}

\institute{
Sterrekundig Instituut Utrecht, Postbus 80000, 3508 TA Utrecht, The
Netherlands.
\and
Centre for Astrophysics and Supercomputing, Swinburne University of
Technology, Hawthorn VIC 3122, Australia 
}

\abstract{
Stellar collisions are 
an important formation channel for blue straggler stars in globular and old
open clusters. Hydrodynamical simulations have shown that the remnants
of such collisions are out of thermal equilibrium, are not strongly
mixed and can rotate very rapidly.
Detailed evolution models of collision products are needed to
interpret observed blue straggler populations and to use them to probe
the dynamical history of a star cluster. We expand on previous studies
by presenting an efficient procedure to import the results of detailed
collision simulations into a fully implicit stellar evolution code. Our
code is able to evolve stellar collision products in a fairly robust
manner and allows for a systematic study of their evolution. 

Using our code we have constructed detailed models of the
collisional blue stragglers produced in the $N$-body simulation of M67
performed by Hurley \emph{et al.} in 2005.
We assume the collisions are head-on and thus ignore the effects of
rotation in this paper.
Our detailed models are more luminous than normal stars of the same
mass and in the same stage of evolution, but cooler than homogeneously
mixed versions of the collision products. The increased luminosity
and inefficient mixing decrease the remaining main-sequence lifetimes
of the collision products, which are much shorter than predicted by
the simple prescription commonly used in $N$-body simulations.
}

\bibpunct{(}{)}{;}{a}{}{,} 

\newlength{\timeswidth}
\newlength{\pluswidth}

\newcommand{\downtriangledot}{\ensuremath{%
\settowidth{\timeswidth}{$\triangledown$}%
\settowidth{\pluswidth}{$\cdot$}%
\addtolength{\timeswidth}{\pluswidth}%
\triangledown\hspace{-0.5\timeswidth}\cdot\hspace{\pluswidth}%
}}

\begin{document}

\maketitle

\section{Introduction}
\label{sec:intro}

Blue stragglers are stars that appear above and blueward of the main
sequence turnoff in the colour-magnitude diagrams (CMDs) of star
clusters, with masses larger than that of a turnoff star. They were first
identified by \citet{sandage_m3} in the globular cluster M3 and soon
afterwards in the old open cluster M67 \citep{1955ApJ...121..616J}.
Blue stragglers have since been found in other globular clusters and
in open clusters of all ages \citep{1995A&AS..109..375A}.  Various
mechanisms have been proposed for their formation
\citep{1993PASP..105.1081S}. Currently, there are 
three
accepted
formation channels: mass transfer due to Roche-lobe overflow in binary
systems, and stellar mergers, either due to dynamical collisions or
through coalescence of close binaries.


\citet{1976ApL....17...87H} first showed that physical collisions
between main-sequence stars are likely to occur in the dense cores of
some globular clusters. In most environments collisions between single
stars are very rare, but binary systems can significantly enhance the
rate of stellar collisions because of their much larger collisional cross
sections. Stellar collisions are thus an important formation channel for
blue stragglers even in old open clusters such as M67, as is
demonstrated by direct $N$-body calculations
\citep{article:hurley_first_m67,article:hurley_m67}.
These simulations indicate that, while all the above-mentioned
formation mechanisms operate in different regions of the cluster and
all are needed to reproduce the observed blue straggler population,
all formation paths -- including binary mass transfer and binary
coalescence -- are strongly affected by the dynamical evolution of the
cluster.  The blue straggler population therefore contains important
information on the dynamical history of a cluster. Extracting this
information requires understanding not only how blue stragglers are
formed but also how they subsequently evolve.


In $N$-body models collisions between two main-sequence stars are
usually approximated by assuming that the stars merge without mass
loss and replacing the end product by a normal (evolved) main-sequence
star with its age reset in accordance with the assumption of complete
mixing \citep{article:tout_evolution_models, article:hurley_bse}.
Smoothed particle hydrodynamics (SPH) simulations of stellar mergers
\citep{1995ApJ...445L.117L,1996ApJ...468..797L} show that some mass is
lost during the collision and that the collision products are
generally far from being fully mixed. To understand how this affects
their further evolution and the predicted blue straggler population,
detailed stellar evolution models of the collision products are
needed.


In early attempts to model the evolution of stellar merger remnants
simplifying assumptions were made: the two stars were assumed to mix
completely \citep{article:bailyn_pinsonneault}, a specified chemical
profile was imposed on the remnant \citep{1995ApJ...455L.163P} or a
simplified prescription for heating during the collision was applied
to mimic the presumed pre-main sequence contraction phase of evolution
\citep{article:sandquist_bolte_hernquist}.  The first realistic
evolution calculations of collision products were performed by
\citet{article:sills_evcolprod1} who took SPH simulations of head-on
collisions between detailed stellar models and imported the SPH
results into a stellar evolution code. Their models demonstrated that
none of the previously made simplifications are valid: although the
collision products are inflated due to shock heating, they do not
develop substantial convection zones during thermal relaxation and do
not undergo significant mixing during their evolution.

This situation changes when the angular momentum of the collisions is
considered. For collisions that are even slightly off-axis, the
remnants retain too much angular momentum to relax into thermal
equilibrium without reaching break-up velocity
\citep{1996ApJ...468..797L}.  The evolution of such collisions was
studied by \citet{sills_evcolprod2}. In the absence of a clear
mechanism by which the stars can lose their excess angular momentum,
they artificially removed a large fraction of the angular momentum to
allow thermal relaxation. Nevertheless, the remnants continue to
rotate rapidly throughout their main-sequence evolution and
rotational mixing makes the remnants bluer and brighter and
significantly extends their lifetimes. These conclusions were
confirmed by higher-resolution calculations
\citep{2002MNRAS.332...49S}. 

The above studies have focussed on blue stragglers in globular
clusters, and investigated only a few interesting cases. This
limitation was imposed by the computation time required and numerical
difficulties in the evolution calculations.  However, the importance
of stellar collisions for the evolution of star clusters calls for a
more systematic approach, covering a larger parameter space and
extending to higher masses and younger clusters. This is the aim of
our current work. We have developed a flexible evolution code that is
able to evolve stellar collision products under a wide range of
circumstances in a fairly robust manner. As our code cannot yet treat
rotation and rotational mixing properly, for the moment we consider
only head-on collisions and ignore the effects of rotation.

As a first step in a systematic study of stellar merger remnants we
have constructed detailed models of the blue stragglers formed by
stellar collisions in the $N$-body model of M67 of
\cite{article:hurley_m67}. They evolved a cluster of 36\,000 stars
from zero age to the age of M67 (4~Gyr) taking into account both
cluster dynamics and stellar and binary evolution. In their simulation
the cluster evolution resulted in 20 blue stragglers at 4~Gyr, eight
of which had a collisional origin. They formed either as a result of
dynamical perturbation of a primordial binary, or as a result of
three-body (binary-single star) or four-body (binary-binary)
interactions. In two of the latter cases, three stars merged in
subsequent collisions with the fourth star ending up as a binary
companion to the blue straggler. Hence in total ten collisions were
involved in the formation of these eight blue stragglers.

%

We evolve these collision products with our detailed stellar evolution
code and compare these models with the evolution tracks of normal
detailed stellar models and fully mixed detailed models, as well as
with the parametric models used by \cite{article:hurley_m67}.  In
particular we investigate the effect on the main-sequence lifetime of
the merger product (\emph{i.e.}, the time during which it will be
visible as a blue straggler),
its position in the Hertzsprung-Russell diagram and the effect on the
chemical abundances of the remnant.
In a companion paper \citep{GlebbeekPols2008} (\refereebf{paper II}) we
study the influence of varying the collision parameters, in particular the
masses of the two stars and their evolutionary stage.


\section{Tools} \label{sec:tools}
\subsection{Modelling the merging process}
To calculate the structure of the collision remnants immediately after the
collision we have used the Make Me A Star (MMAS) code developed by
\citet{article:lombardi_mmas}, which produces a one-dimensional model that 
approximates the results of detailed smooth particle hydrodynamics (SPH)
calculations.
The essence of the approximation is the observation that in a stellar model
in hydrostatic equilibrium the quantity $A$ defined by
\begin{equation}\label{eqn:defa}
A^{3/2} = \frac{(2\pi\hbar^2)^{3/2}}{m_\mathrm{u}^4} 
   \frac{1}{\mu^4 }
   \prod_i\left[
      \left(\frac{B_i}{\mu}\right)^{5/2}\frac{\omega_i}{X_i}
   \right]^{-Y_i}
\mathrm{e}^{S - \frac{5}{2}}
\end{equation}
increases monotonically from the centre to the surface, at least as long
as radiation pressure is negligible 
\refereebf{\citep[See for instance][for details]{1996ApJ...468..797L,article:lombardi_mmas}}
Here, $S$ denotes the entropy per particle 
in units of Boltzmann's constant, $\mu$ is the mean molecular weight of the 
constituent particles, $B_i$ is the mass of particle species $i$ in units of 
the atomic mass $m_\mathrm{u}$, $X_i$ and $Y_i$ are their abundance fraction 
by mass and number, respectively. Note that $A$ is a function of entropy and
composition only.
The factor $\omega_i$ denotes the degeneracy of the ground state and is taken 
as $1$ for nuclei and $2$ for electrons.
For a classical ideal gas, (\ref{eqn:defa}) reduces to
\begin{equation}
A = \frac{P}{\rho^{5/3}}.
\end{equation}

MMAS modifies the initial $A$ profiles of the parent stars to correct
for shock heating during the collision and estimates the amount of mass lost.
The amounts of shock heating and mass loss are calibrated to detailed 
SPH results.
The remnant profile is then constructed by collecting all mass-bins from the
parent stars, sorting them in order of increasing $A$ and integrating the
equations of hydrostatic equilibrium and mass conservation.

The composition profile of the remnant is also determined from the $A$
profile. Shock heating leads to partial mixing of adjacent layers in
each of the parent stars. After both $A$ and the individual
composition profiles have been corrected for shock heating, the
composition in the remnant is set to the average composition of
material from the parent stars with that value of $A$.
So if for a particular value of $A$ a fraction $f_1$
comes from the primary with composition $X_1$ and a fraction $f_2$ comes
from the secondary with composition $X_2$, then the composition
$X_\mathrm{r}$ of the corresponding layer in the remnant is
\beq
X_\mathrm{r} = X_1 f_1 + X_2 f_2.
\eeq
For more details see \citet{article:lombardi_mmas}.

\subsection{The stellar evolution code}\label{sec:stevcode}
To calculate the evolution of the parent stars as well as the 
further evolution of the collision product we use
the stellar evolution code (hereafter STARS) originally developed by Eggleton
\citep{article:eggleton_evlowmass} and updated by others
\citep{article:pols_approxphys}.
The code uses an adaptive non-Lagrangian mesh that allocates meshpoints
according to a mesh-spacing function that places more meshpoints in regions
where a higher resolution is required.
This means that stars can be evolved with reasonable accuracy with as few as
$200$ meshpoints.

Our version of STARS uses nuclear reaction rates from \citet{article:caughlan_and_fowler_reactionrates}
and \citet{article:caughlan_reactionrates}
and opacities from \citet{article:opal1992} and \citet{article:alexander_ferguson_lowtopac}. The assumed heavy-element composition is scaled to solar
abundances \citep{article:anders_and_grevesse_abund}.

STARS is fully implicit and solves the equations for the structure and 
composition of the star simultaneously. Convection is treated using the
mixing-length prescription
\citep{article:bohm-vitense_convection} and convective mixing is modelled as a 
diffusion process \citep{article:eggleton_mixingproc}. 
We use a ratio of mixing length to local pressure scale height $l/H_P=2.0$. 

Usually the mean molecular weight in stars will be either constant in a region or
decrease radially outward. In merger remnants there can be layers in which the
molecular weight gradient is inverted and a layer of higher mean molecular
weight lies on top of a layer of lower mean molecular weight. Such a situation
is unstable and leads to a process known as thermohaline mixing
\citep{article:ulrich_thermohaline, article:kippenhahn_thermohalinemixing}. 
\refereebf{The instability arises because material with high molecular
weight is supported by thermal buoyancy. When a fluid element exchanges
heat with its environment it loses buoyancy and begins to sink on the
timescale for heat exchange, \emph{i.e.} the local thermal timescale.}
We model this as a diffusion process as described in 
\citet{article:stancliffe_thmixing}.
\refereebf{The diffusion coefficient is given by the product of the typical
velocity and size of the fluid elements and the efficiency of mixing thus
depends on their assumed geometry. The efficiency we have adopted in
this work corresponds to spherical geometry, as in
\citet{article:kippenhahn_thermohalinemixing}.
If the fluid blobs are elongated the mixing is more efficient. However,
since for our adopted choice the mixing occurs on the local thermal
timescale and is fast compared to the nuclear timescale, our results are not
sensitive to an increase in the efficiency of thermohaline mixing.}

STARS uses a simple model for convective overshooting \citep{article:schroeder_overshooting}
that allows extra mixing in regions where $\nabla_\mathrm{rad} - \nabla_\mathrm{ad} > -\delta$.
We have found that this prescription leads to spurious mixing in the cores of 
some of our collision products (see \S \ref{sec:overshooting}), so we have 
chosen to disable convective overshooting for this work.

\subsection{Constructing starting models for merger remnant evolution}

To construct starting models for the collisions we evolved stars of the
appropriate masses from the zero-age main sequence (ZAMS) to the time of 
collision as listed in Table~\ref{tab:collisions} 
(see \S \ref{sec:evolution}). 
These models were then used as input for MMAS.

\begin{figure}
\includegraphics[width=0.5\textwidth]{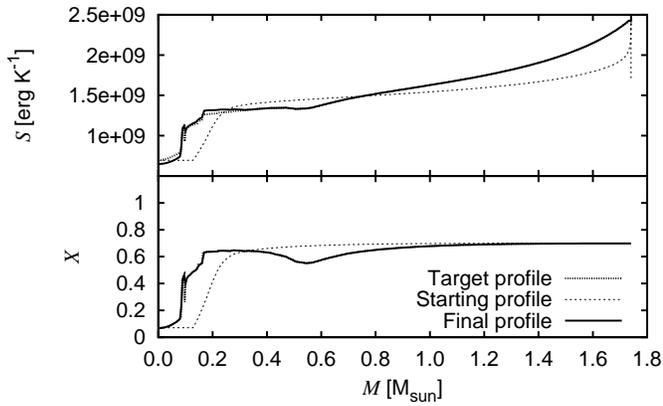}
\caption{Entropy (top panel) and hydrogen abundance (lower panel) profiles
in the merger remnant 2321 (see Table~\ref{tab:collisions}) as a
function of mass coordinate during the convergence stage. The profile
at the start of the iteration is indicated by a thin dotted line while
the target profile is indicated by a thick dotted line.  The solid
line is the final profile after the import is complete.}
\label{fig:convergence_monitor}
\end{figure}

\begin{table*}
\caption{Initial conditions for the collisions and collision product lifetimes.
The first column lists the number by which the star is refered to in
\citet{article:hurley_m67}, followed by the time of collision $t$, the masses
$M_1$ of the primary and $M_2$ of the secondary, the total mass $M$ of the
remnant according to the detailed models and the remaining main sequence lifetime
$t_\mathrm{ms}$. For comparison we also list the main sequence lifetime
$\tau_\mathrm{ms}$ of a normal star of mass $M$, the lifetime 
$\tau_\mathrm{ms,hom}$ of a homogenised star and the remaining lifetime of the merger
remnant according to the parametric BSE prescription ($t_\mathrm{ms,BSE}$).
Time is given in millions of years (Myr) and masses are given in solar masses
($\mathrm{M}_\odot$).
}\label{tab:collisions}
\begin{center}
\begin{tabular}{lllllrrrr}
\hline\hline
ID & $t$ & $M_1$ & $M_2$ & $M$ & $t_\mathrm{ms}$ & $\tau_\mathrm{ms}$ & $\tau_\mathrm{ms,hom}$ & $t_\mathrm{ms,BSE}$\\
\hline
  2203 & 3480 & 1.23 & 0.85 & 1.93 &  539 & 1000 &  745 &  991 \\ 
  2321 & 3960 & 1.29 & 0.59 & 1.74 &  599 & 1345 &  862 & 1305 \\ 
  2565 & 3650 & 0.95 & 0.94 & 1.76 & 1016 & 1297 & 1122 & 1336 \\ 
  2973 & 3170 & 0.89 & 0.80 & 1.56 & 1569 & 1850 & 1685 & 1877 \\ 
  3121 & 3890 & 1.09 & 0.54 & 1.52 & 1428 & 1993 & 1629 & 2051 \\ 
3289-1 & 3610 & 0.82 & 0.51 & 1.25 & 3157 & 3736 & 3736 & 3910 \\ 
3289-2 & 3610 &      & 0.76 & 1.86 & 1024 & 1103 & 1058 & 1020 \\ 
  3445 & 2770 & 0.77 & 0.76 & 1.41 & 2172 & 2502 & 2361 & 2547 \\ 
3835-1 & 3797 & 0.82 & 0.60 & 1.32 & 2657 & 3090 & 3090 & 3191 \\ 
3835-2 & 3798 &      & 0.31 & 1.56 & 1765 & 1849 & 1793 & 1776 \\ 
\hline
\end{tabular}
\end{center}
\end{table*}

The output from MMAS was converted into an input model for STARS
using implicit calculations starting from a normal ZAMS model with the same
mass as the collision remnant. This model is first evolved until its core 
hydrogen abundance matches that of the remnant. At this point, the entropy and
composition profiles of the model are adjusted to reproduce the profiles of 
the collision product. Figure \ref{fig:convergence_monitor} shows the profiles
at the beginning and end of the iteration as well as the output profiles from
MMAS for a generic case.
The entropy profile was adjusted by adding an artificial energy source 
$\epsilon_\mathrm{art}$ to the luminosity equation,
\beq
\diff{L}{}{m} = \epsilon_\mathrm{nuc}
-T \diff{S}{}{t}
+\epsilon_\mathrm{art}.
\eeq
Here, $\epsilon_\mathrm{nuc}$ is the net energy generation rate from nuclear
reactions, $S$ is entropy per unit mass, $T$ is the temperature and
$-T\mathrm{d}S/\mathrm{d}t$ is the energy released by gravitational contraction.
This term vanishes for stars in thermal equilibrium. As mentioned collision
remnants are out of thermal equilibrium and the effect of 
$\epsilon_\mathrm{art}$ is to specify $T\mathrm{d}S/\mathrm{d}t$. A suitable
form is given by
\beq
-T \diff{S}{}{t}+\epsilon_\mathrm{art} = T \frac{\Delta S}{\tau},
\eeq
where $\Delta S$ is the difference in entropy between the current model and the
target model and $\tau$ is an artificial timescale on which the adjustment is
to be made. With this choice an equilibrium is reached when the entropy
profile in the model matches that of the target model, $\Delta S = 0$. The
timescale $\tau$ is arbitrary in principle and can be chosen to change the
relative weight of the terms in the energy equation, which affects the speed of
convergence. Choosing $\tau$ to be of the order of the current timestep was
found to work well.

At the same time, the composition profile is changed smoothly 
by setting
\beq
X \to (1 - \lambda) X + \lambda X_\mathrm{target},
\eeq
where $\lambda$ is increased monotonically from $0$ to $1$ in the course of
the run. Composition changes due to nuclear reactions and mixing processes were
ignored. We continue this procedure until the entropy and composition profiles
in the model match those of the collision product. The final model is then 
used as a starting model for the evolution of the merger remnant.

This implicit scheme is very stable and flexible, and deals well with small
irregularities in the output. 
Figure \ref{fig:convergence_monitor} shows that the resulting entropy and
composition profiles agree very well with the target profiles. Except
for the entropy profile in the core the two curves overlap within the
thickness of the lines shown. We have tested whether small deviations like
these affect the long-term evolution of the collision product and have
found that they are unimportant.

Detailed SPH calculations do not have sufficient resolution to resolve the
outer parts of the envelope of the collision product and MMAS likewise does
not have any real information about this region. This means that the structure 
of the outer envelope cannot be determined from these models.
\citet{article:sills_evcolprod1} extrapolated the entropy profile and used
the condition of hydrostatic equilibrium to reconstruct the outer layers.
We have found it easier to assume that these layers are in thermal equilibrium
and have not tried to impose a particular entropy profile in these layers.
This is reasonable considering that the local thermal timescale is short
compared to that of the rest of the star. 
\refereebf{Because we do not enforce a large entropy on the outer
layers our models are somewhat less inflated than those of
\citet{article:sills_evcolprod1} and consequently start at a lower
luminosity.}
As pointed out by \citet{article:sills_evcolprod1}
the long term evolution of the collision product is determined by the interior
properties and is not sensitive to the assumptions made for the outer layers.
\refereebf{Comparing their evolution tracks of collision products to tracks
computed with our code for the same masses and metallicity confirms this.
We find that the contraction timescales are similar and that the tracks
agree from the main sequence onward.}

In this work we ignore rotation in the collision products by assuming
that all collisions are head-on. In the more realistic case of
off-centre collisions the effect of rotation on the evolution of
collision products can be substantial, as outlined in
\S\ref{sec:intro}.  We choose to ignore this problem for the moment
and we defer a discussion of this limitation until \S\ref{sec:rotation}.


\subsection{The BSE/NBODY4 prescription}

We will compare the outcome of our detailed evolution models with the
results obtained in the $N$-body calculations of
\citet{article:hurley_m67}.  These calculation were performed using the
NBODY4 code \citet{1999PASP..111.1333A} in which binary evolution is
provided by the Binary Star Evolution (BSE) algorithm
\citep{article:hurley_bse}. In this algorithm a simple analytic
prescription is used to model the outcome of stellar collisions where no
mass is lost during the collision, i.e.\ the remnant mass is $M_1+M_2$. The
merger remnant is replaced by a normal evolved main-sequence star with a
starting age $t'$ given by
\beq \label{eq:age_bse}
t' = \frac{1}{10} \frac{\tau_\mathrm{ms}}{M_1+M_2} \left(
M_1 \frac{t}{\tau_\mathrm{ms,1}} + 
M_2 \frac{t}{\tau_\mathrm{ms,2}} 
\right),
\eeq
so that the remaining lifetime of the collision product is
$t_\mathrm{ms,BSE} = \tau_\mathrm{ms} - t'$. Eq.~(\ref{eq:age_bse}) is
based on the implicit assumption that the remnant is fully mixed. Here
$\tau_\mathrm{ms,1}$, $\tau_\mathrm{ms,1}$ and $\tau_\mathrm{ms}$ are
the main-sequence lifetimes of the two colliding stars and of a normal
star with mass $M_1+M_2$, respectively. These lifetimes are calculated
according to the analytic formulae of
\citet{article:hurley_sse} which are based on detailed models that
include convective overshooting \citep{article:pols_evmodels}.

\section{Evolution of the merger remnants} \label{sec:evolution}
\begin{figure*}
\includegraphics[width=\textwidth]{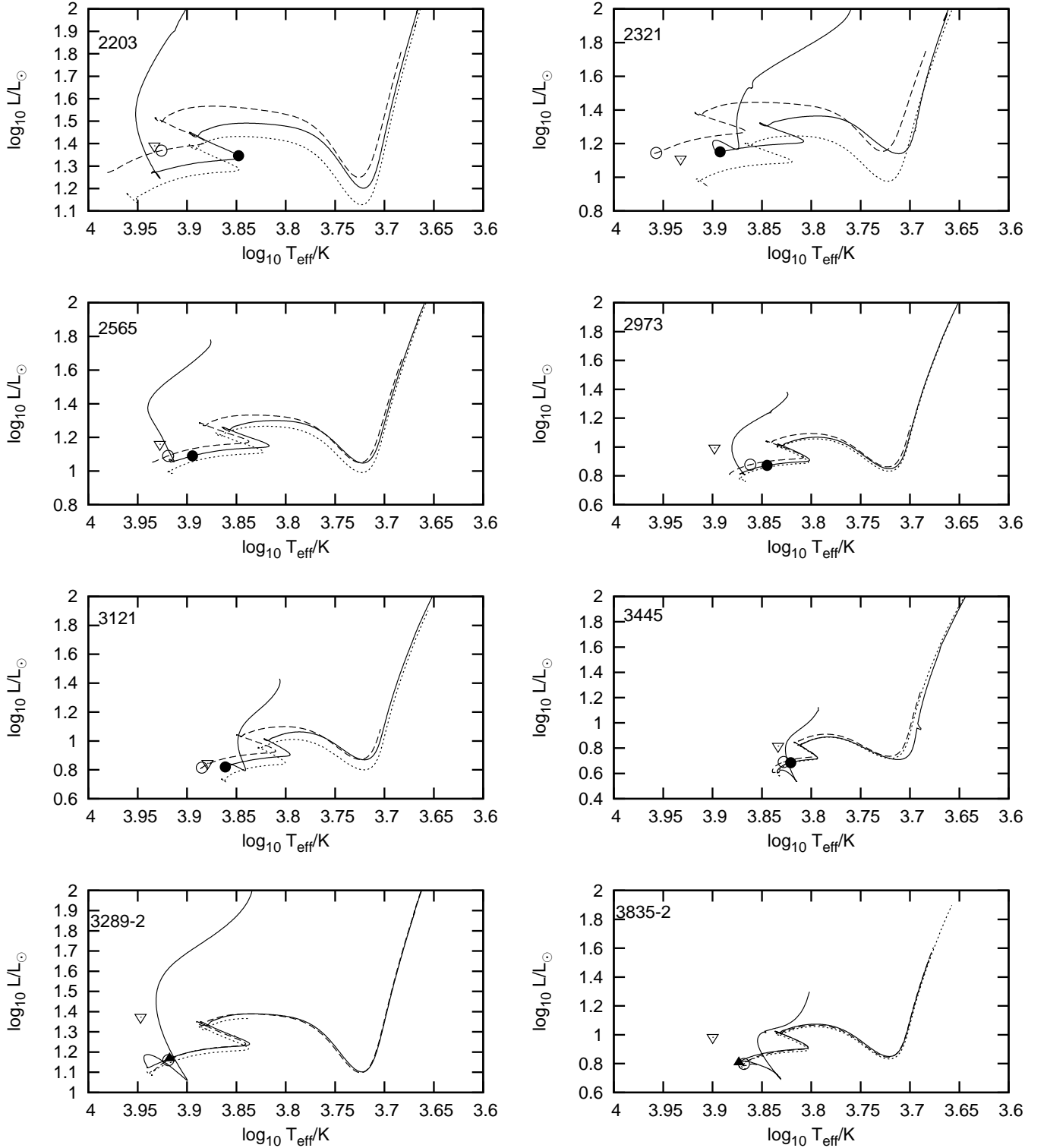}
\caption{Evolution tracks for the collision products (solid lines) compared to tracks
for homogenised models (dashed lines) and a main sequence star of the same ZAMS mass (dotted line).
Also marked are the positions at $4 \, \mathrm{Gyr}$ for the detailed models 
($\bullet$ for remnants of a single collision and $\blacktriangle$ for the 
remnants of two collisions), the homogenised
models ($\odot$) and the BSE prescription ($\downtriangledot$).}
\label{fig:all_hrd}
\end{figure*}

The masses $M_1$ and $M_2$ of the colliding stars and the time of
collision $t$ are taken from the $N$-body simulation of M67 by
\citet{article:hurley_m67} and are listed in Table
\ref{tab:collisions}. The collisions listed as 3289-2
and 3835-2 are further collisions between a collision product and another
main sequence star. In the case of 3289-2 this results in a collision
product that is more than twice as massive as the most massive progenitor
star. The most massive collision product in our list, however, is 2203.

We have constructed starting models for the calculation of the evolution of the
collision products in Table \ref{tab:collisions} using the method described 
in \S \ref{sec:tools}. All models are calculated for an initial hydrogen 
mass fraction $X=0.70$ and a mass fraction of heavy elements $Z=0.02$.
For the double collisions 3289 and 3835 we have used the output of the
first collision as input for the second collision. All collision products
were then allowed to evolve until the tip of the giant branch.

For each evolution model based on the MMAS output we have calculated
two evolution tracks for comparison. One is a star with the same mass
as the collision product that is evolved from the ZAMS at composition
$X=0.70$ and $Z=0.02$, the other is a homogenised version of the
collision product.

Table~\ref{tab:collisions} also lists the total remnant mass $M$
according to MMAS, the remaining main-sequence lifetime $t_\mathrm{ms}$ 
according to our detailed evolution models, the main-sequence lifetime 
$\tau_\mathrm{ms}$ of a star with the same ZAMS mass and the lifetime 
of a homogenised version of the collision product, $\tau_\mathrm{ms,hom}$. 
The lifetimes listed for 3289-1 and 3835-1 are the lifetimes these merger
remnants would have had if they had not been involved in a second
collision.
Also given in Table~\ref{tab:collisions} is the remaining lifetime 
according to the BSE prescription, $t_\mathrm{ms,BSE}$ 
(see \S\ref{sec:compare} for a comparison and discussion). 

In Figure \ref{fig:all_hrd} we have plotted the evolution tracks of our models
in the Hertzsprung-Russell diagram. The solid lines are the evolution tracks
of the detailed collision products while the dotted lines are the evolution tracks of
the normal stars. The homogenised models are indicated by a dashed line. We see
that in general, the homogenised models are hotter and brighter than the normal
models, while the collision products tend to be brighter than the normal stars
but less luminous than the homogenised models. This is an opacity effect,
as will be discussed below.


\subsection{Initial structure and contraction phase}

Initially, all collision products have a large amount of excess thermal energy
and the star's main energy source is gravitational contraction. We refer to
this initial phase where the collision product is puffed up and out of thermal
equilibrium as the contraction phase.

There is no significant mixing during the collision. This can lead to composition
inversions in the remnant: the core is rich in helium, on top of which there is
a hydrogen-rich layer above which there is again a helium-rich layer of 
material from the core of the primary. These composition inversions show up in 
models  where the primary star is sufficiently evolved to have burned a
significant fraction of its central hydrogen to helium while the secondary is 
relatively unevolved, \emph{i.e.} where the secondary is much less massive than
the primary.
Consequently, they are present in models 2203, 2321 and 3121 as well as the
two double mergers 3289 and 3835, although the composition inversions are
small in this case.
As mentioned in section \ref{sec:stevcode}, a composition inversion of this
type is secularly unstable and leads to thermohaline mixing.
This homogenises part of the central region of the remnant.

\begin{figure}
\includegraphics[width=0.5\textwidth]{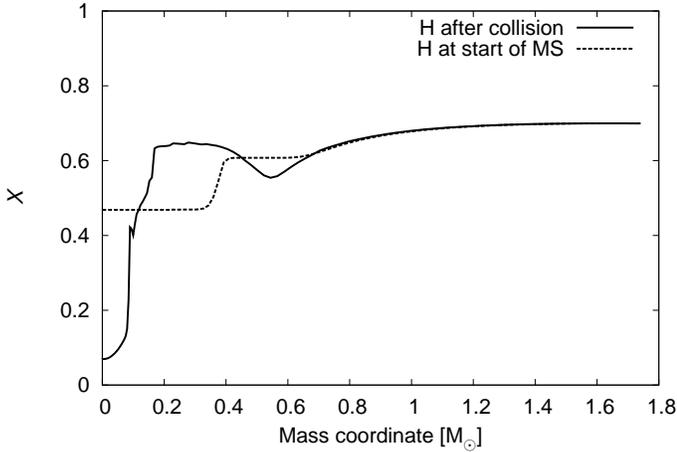}
\caption{Composition profiles in the merger remnant 2321, immediately 
after the collision (solid line)
and on reaching the main sequence (dashed line).}
\label{fig:2321_comp}
\end{figure}

As an illustrative example, we will discuss in more detail the various evolutionary stages
of the collision product 2321. This exhibits
many of the transient features that occur during the evolution of these
collision products.

First we consider the composition profile for collision $2321$ 
which is shown in Figure 
\ref{fig:2321_comp}. 
During the collision a large portion of the core of the $1.29 M_\odot$ primary 
has sunk to the  centre, creating a hydrogen poor core below mass coordinate 
$0.1 M_\odot$. Above this region material is slightly mixed with material from
the $0.59 M_\odot$ secondary, leading to a hydrogen-rich layer between $0.2$
and $0.4 M_\odot$. On top of this there is a helium-rich layer at mass coordinate
$0.6 M_\odot$. Between this layer and the surface of the star the composition 
tends to the primordial composition.

The helium-rich layer at $0.6 M_\odot$ is unstable to thermohaline mixing, 
which will mix the helium inward while the star contracts to the main sequence.
At the same time, hydrogen will reignite in a shell at $0.1 M_\odot$. 
This burning shell forms as a result of a peak in the hydrogen burning rate due to the
steep increase of the hydrogen abundance in this region.
The shell drives a convection zone that connects to the thermohaline layer and
mixes the inner $0.4 M_\odot$ of the star. This mixes helium-rich material into
this burning shell which has the effect of lowering its efficiency. By the time
the merger remnant has reached the main sequence the burning shell has
extinguished. 
We define the end-point of the contraction phase as the moment where central
hydrogen burning takes over as the main energy source and the star
is in thermal equilibrium.

At this point the central hydrogen abundance has increased to a 
mass fraction of $0.46$ and the composition profile has changed to the dashed 
line in Figure \ref{fig:2321_comp}. The collision product has a convective
core on the main sequence that extends to about $0.2 M_\odot$.

The evolution of the core is shown in Figure \ref{fig:rhoc_tc}, where the
central temperature $T_\mathrm{c}$ is plotted against the central density
$\rho_\mathrm{c}$. The short-dashed lines indicate lines of constant entropy 
with entropy increasing in the direction of increasing temperature and 
decreasing density. 
The core starts at point $a$ with an entropy that is close to that of
the core of the former primary. This means that the core has an entropy that is
too low for a star of its mass and it will need to increase its entropy before
the star can come into thermal equilibrium. This increase in entropy is 
achieved by
expansion, which means that the core moves to the left in the diagram until it
reaches the correct adiabat. At this point the core starts to contract again
until thermal equilibrium is finally reached at point $b$.
At this point the core is slightly hotter than the core of a star that was 
born with the same ZAMS mass.

\subsection{Main sequence evolution}

\begin{figure}
\includegraphics[width=0.5\textwidth]{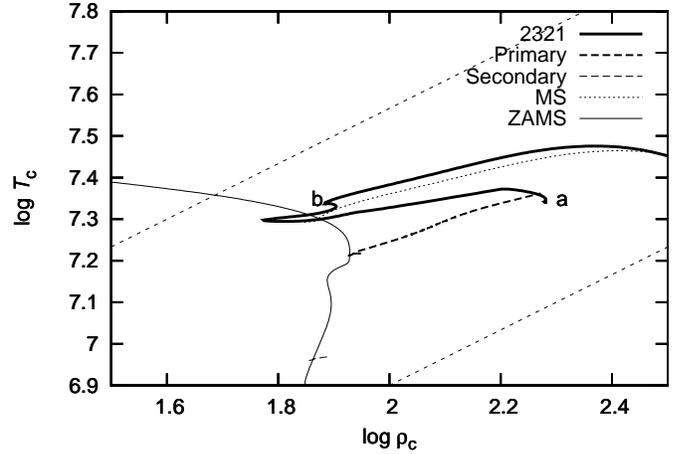}
\caption{Evolution track for the collision product 2321 in a 
$\rho_\mathrm{c}$--$T_\mathrm{c}$ diagram (thick solid line). Also
plotted are the progenitor tracks (dashed lines), the track for a star
with the same ZAMS mass as the collision product (dotted line) and the
location of the ZAMS (thin solid line).  The short-dashed lines
represent lines of constant entropy for a given composition.  The
collision product starts with an overdense core at the point labeled
$a$ and reaches the main sequence at point $b$.}
\label{fig:rhoc_tc}
\end{figure}


After the contraction phase the collision products follow more or less normal 
main-sequence tracks, although their luminosity (Figure~\ref{fig:all_hrd}) and 
central temperature (Figure~\ref{fig:rhoc_tc}) are typically higher than they 
would have been for a main sequence star of the same ZAMS mass. This is owing to 
the helium enhancement in their envelopes, which increases the mean molecular
weight $\mu$. The luminosity $L_\mathrm{merger}$ of the collision product 
scales with the luminosity of a normal star according to the homology relation 
\citep{book:kippenhahn_weigert}
\beq\label{eqn:homology_luminosity}
L_\mathrm{merger} \approx L_\mathrm{ms}
\left(\frac{\mu_\mathrm{merger}}{\mu_\mathrm{ms}}\right)^4,
\eeq
with $\mu_\mathrm{merger}$ and $\mu_\mathrm{ms}$ the mass-averaged mean molecular
weight in the collision product and the main sequence star when these have
the same effective temperature.
The homology relation (\ref{eqn:homology_luminosity}) is strictly valid only
for homogeneous stars with constant opacity but it reproduces the luminosity
shift of the merger remnants with respect to normal main sequence stars very
well.

The higher effective temperature of the homogenised models results from a
reduction of the average opacity owing to the increased helium content of their
envelopes.
The opacity affects the stellar structure by increasing or decreasing
the photon mean free path. In the weakly mixed collision products,
most of the helium enhancement is in the compact interior while the
more extended envelope has the normal ZAMS composition. Thus,
throughout most of the volume of the star a photon will `see' a normal
hydrogen rich composition and the opacity is not strongly
affected. Conversely, in the homogenised models the helium enhancement
is present in the entire envelope and a photon will `see' a
helium-enhanced composition with a lower opacity, leading to a more
compact structure.


Because the luminosity is enhanced the collision products do not lie 
exactly on the extension of the main sequence, but can lie somewhat above it.
As a consequence of their higher luminosity, the central temperature of
the collision products is slightly increased (see Fig.~\ref{fig:rhoc_tc}) 
and the main-sequence lifetime (\emph{i.e.} the time until core hydrogen 
exhaustion) is reduced compared to the lifetime of a normal
main sequence star with a similar composition in the core.

\subsection{Hertzsprung gap and first giant branch}
\begin{figure}
\includegraphics[width=0.5\textwidth]{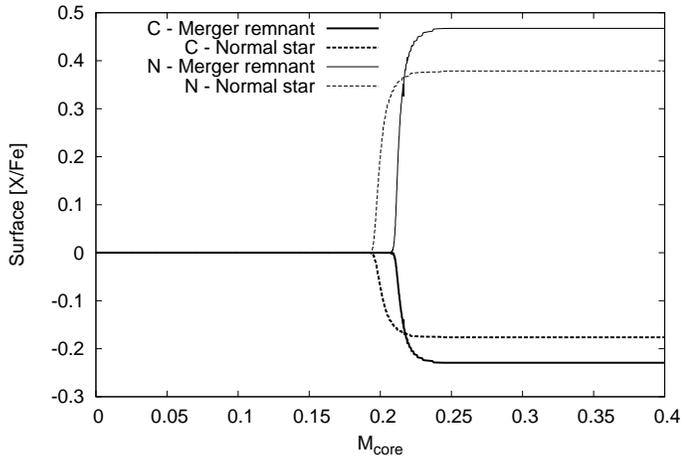}
\caption{Surface [C/Fe] and [N/Fe] abundances as a function of the helium core
mass (as a measure of age) for collision product 2321. 
The carbon abundance (thick solid line) in the collision product is depleted 
when compared to a normal red giant (thick dashed line) while the nitrogen
abundance (thin lines) is correspondingly enhanced.}
\label{fig:scomp_cno}
\end{figure}

We follow the evolution of the collision products through the
Hertzsprung gap up to the tip of the giant branch. We were unable to calculate
models beyond the helium flash.

The red giant phase is very similar to that of a normal star. 
The first dredge-up occurs when the convective envelope penetrates into 
the helium-rich layer. For our example collision product $2321$ this 
increases the surface helium abundance to 
$Y_\mathrm{s} = 0.334$ compared to $Y_\mathrm{s} = 0.296$
in a normal red giant of the same mass.
More of the dredged-up material has undergone processing than in a normal red 
giant because part of this material comes from the core of the
original primary. As a result, more carbon has been converted to nitrogen
by CN cycling, which enhances the nitrogen abundance and depletes the carbon
abundance compared to a normal red giant, both by $\sim 0.1$ dex 
in the case of collision 2321 as shown in Figure \ref{fig:scomp_cno}.
\refereebf{Unfortunately this does not exceed the typical observational
error bar of $0.15$ dex \citep[for instance][]{2000A&A...354..169G}.}

\subsection{Double collisions}
In the $N$-body simulation two blue stragglers are present at $t = 4$~Gyr
that resulted from consecutive collisions between three stars \refereebf{in
a binary-binary interaction}. 
In both cases the second collision happened \refereebf{almost immediately
after the first. We have not considered the situation where the time
between collisions is long enough for the collision product to have evolved
before the second collision}.
We have calculated the outcome of the first collision and used that 
as input for the second collision. 
%
These double collision remnants evolved in a similar way to single collision 
products and their main sequence evolution is very similar to that of ordinary 
stars of their mass.
This can be understood from the fact that their progenitors are all fairly 
low-mass stars that are essentially unevolved at the time of collision. 
For this reason collision product 3289-2 has the highest effective temperature
(Fig.~\ref{fig:all_hrd}) and the bluest colour (see Fig~\ref{fig:hrd} 
in \S\ref{sec:compare}), even though it is not the most massive collision 
product.
It is likely that the structure and evolution of double collisions involving 
more massive and thus more evolved stars, or with more time passing between 
the collisions, will produce remnants that stand out more. The bluest blue
straggler in M67 (S977; \citealp{mathys_bss_m67}) is quite possibly the result 
of a double collision.

\section{Comparison of different methods} \label{sec:compare}
\begin{figure}
\includegraphics[width=0.5\textwidth]{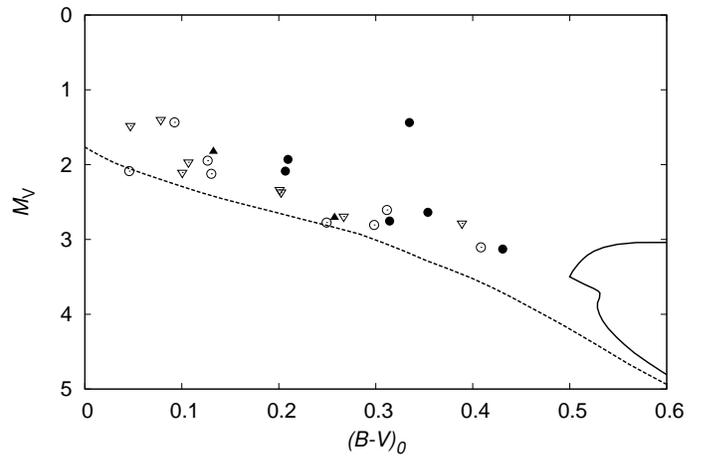}
\caption{Location in the colour-magnitude diagram of the blue stragglers at $4 \mathrm{Gyr}$.
Detailed models of the single collision remnants are indicated by $\bullet$ 
and double collision remnants by $\blacktriangle$. Homogenised models are 
indicated by $\odot$ and the BSE prescription is indicated by
$\downtriangledot$.
The isochrones for $t=0$ and $t=4\mathrm{Gyr}$ are constructed from the models
by \citet{article:pols_evmodels}.
}
\label{fig:hrd}
\end{figure}
In Fig.~\ref{fig:all_hrd} we show as solid symbols the locations 
in the Hertz\-sprung-Russell diagram of each of the merger products listed in 
Table~\ref{tab:collisions} after $4\,\mathrm{Gyr}$ (the age of 
M67 obtained by isochrone fitting in \citealp{article:pols_evmodels}) and
compare these to the locations of the homogenised models (open symbols) 
and the positions according to the BSE prescription (triangles).
In Figure~\ref{fig:hrd} we show the corresponding locations for all collision
products in a colour-magnitude diagram.
\refereebf{A comparison between the observed blue stragglers and our models
is given in paper II.}
We converted the theoretical surface parameters
($L$, $T_\mathrm{eff}$ and $\log g$) to observable magnitude and colour 
($M_V$, $B-V$) using model atmospheres from \citet{article:kurucz} 
with empirical corrections from \citet{article:lejeune_calibration_of_spectra}.
Also shown is the location of the ZAMS (dashed line) and the 
$t=4\,\mathrm{Gyr}$ isochrone from \citet{article:pols_evmodels}.
The detailed models are redder than the homogenised models and the positions
predicted by the BSE recipe, with the largest difference occurring for the
brightest and most massive blue stragglers.

The homogeneous models are bluer than the detailed models for two
reasons. First, the helium abundance in the envelopes of the
homogeneous models is enhanced over that of the detailed models, which
moves the position of the ZAMS line to the blue. Second, if we compare
the lifetimes listed in Table~\ref{tab:collisions} we see that the
remaining lifetime for the homogenised models is longer than that of
the detailed models, so that the homogenised models are closer to
their zero-age main sequence.  This is particulary striking in remnant
2203 (Fig.~\ref{fig:all_hrd}), for which the detailed model has almost
reached the terminal-age main sequence point. The longer lifetime of
the fully mixed models is due to the larger hydrogen abundance in
their cores compared to the detailed models. The largest discrepancy
is found for progenitor stars that were substantially evolved at the
time of collision, in particular remnant 2321. The best match between
detailed and fully mixed models is obtained for the double collision
remnants, whose progenitor stars were not very evolved.  In general
the approximation of homogeneous mixing is worse for collisions
involving more evolved stars than for collisions involving less
evolved stars.

The BSE recipe gives locations that are brighter and bluer mainly
because mass loss from the collision is not considered, leading to
higher remnant masses. As can be seen from Figure~\ref{fig:all_hrd}
this effect is most noticeable in remnants 2973 and 3445 and the two
double collision remnants 3289-2 and 3835-2.  Because BSE simply uses
a normal main-sequence star model for the collision remnant, the
effect of the higher mass on the luminosity is at least partially
compensated by neglecting the effect of helium enhancement.

The BSE prescription assumes complete mixing to estimate the remaining
lifetime of the merger remnant.  The BSE lifetimes are generally
longer than those of both our detailed models (by more than a factor 2
in the case of remnant 2321) and the fully mixed models, despite the
fact that the overestimate of the remnant mass by BSE would result in
an underestimate of the lifetime by 20--30~per cent. This is, however,
counteracted by two effects when compared to our fully mixed
models. First, the neglect of helium enhancement leads to an increase
of the lifetime by up to 30~per cent for collision remnants of evolved
progenitor stars. Second, the inclusion of convective overshooting in
the BSE models gives rise to longer main-sequence lifetimes, by
25--30~per cent in the mass range considered (see \S\ref{sec:rotation}
for a discussion of this issue). A fairer comparison of the BSE
\emph{method} is to compute the remnant lifetime using
eq.~(\ref{eq:age_bse}) but inserting main-sequence lifetimes of
non-overshooting models; we then find BSE lifetimes that are indeed
somewhat smaller than those of our fully mixed models.

The homogeneous models and the BSE prescription can both produce blue 
stragglers 
that are brighter and bluer than those produced by the detailed models. In
the case of the BSE models this is due to the higher mass of the collision
products, while in the case of the homogeneous models this is due to the
higher helium content of their envelopes. This implies that in order to
form the brightest blue stragglers it may be necessary to collide more massive
progenitor stars, or that enhanced mixing in the collision products (for 
instance due to rotation) is necessary.

\section{Discussion}

\subsection{Convective Overshooting} \label{sec:overshooting}

There is evidence that for ordinary stars in the mass range we
consider mixing extends beyond the boundary of the classical
convective core, which is usually modelled by convective overshooting
as discussed in \citet{article:schroeder_overshooting}. As mentioned
in section \ref{sec:stevcode}, we have disabled convective
overshooting for the evolution of the merger remnants.  During the
contraction phase in some of the remnants, regions develop where
$\nabla_\mathrm{rad} - \nabla_\mathrm{ad}$ is close to zero but
remains negative ($\nabla_\mathrm{rad} - \nabla_\mathrm{ad} >
-\delta$), so that these regions never actually become convective.
The overshooting prescription used in our code leads to undesired
spurious mixing in such regions.

Convective overshooting affects the lifetime and the shape of the
main-sequence track of a star.  The progenitor stars will mostly not
be overly affected by overshooting (because their masses are too small
to develop convective cores), whereas the collision product is massive
enough to experience overshooting.  The main difference we expect is
that for two given parent stars the collision product will have a
larger reservoir of hydrogen and live longer. The inclusion of
overshooting in our detailed models would likely make the collision
remnants of evolved parent stars (such as 2203) somewhat bluer,
because the enhanced mixing brings the remnant closer to its ZAMS
position. More importantly, the lifetimes of all our collision
products would be increased, perhaps by about 30~per cent (based on
comparing overhooting and non-overshooting models from
\citealp{article:pols_evmodels}).


\subsection{Rotation} \label{sec:rotation}

We have ignored rotation in the present work by treating all
collisions as head-on. Rotation will modify our results by extra mixing and
possibly by enhanced mass loss. The latter is particularly relevant because
off-centre collisions produce remnants that have so much angular momentum
that their rotation rate will approach critical rotation during the contraction
phase. How the remnant loses angular momentum is still an unsolved problem
and it is likely that magnetic fields play a role here, possibly by coupling 
the star to a circumstellar disk \citep{sills_evcolprod3}. 
At present, nothing is known about the magnetic field in the collision
product after the collision. 

Rotational mixing provides a larger hydrogen reservoir for nuclear
burning and increases the helium content in the envelope, making the
remnants bluer and brighter and extending their lifetimes
\citep{sills_evcolprod2}. By ignoring rotation we implicitly assume
that an efficient spin-down mechanism operates in the collision
products such that they avoid both the angular momentum problem and
significant rotational mixing. It is significant that the blue
stragglers in M67 are slow rotators, with projected rotation
velocities smaller than typical for their spectral type
\citep{1984ApJ...279..237P,mathys_bss_m67}. Since all formation mechanisms
for blue stragglers are expected to result in rapid rotation, this
suggests that they can indeed lose angular momentum efficiently. We
will investigate this problem in more detail in future work.

\section{Conclusions}

Hydrodynamical simulations of stellar collisions produce remnants that
are out of thermal equilibrium and are not fully mixed.  We have
developed an efficient procedure for importing the structural and
chemical profiles of such collision products into a fully implicit
stellar evolution code. Our evolution code is fairly robust and can
evolve the collision remnants until the tip of the giant branch with a
minimum of human intervention. We have applied our code to construct
detailed models of collisional blue stragglers formed in the $N$-body
simulation of M67 by \citet{article:hurley_m67}.

The evolution of collision products depends on the amount of mixing
during the collision and the thermal relaxation phase. Assuming the
collision product has been homogeneously mixed produces evolution
tracks that are too blue while replacing the collision product with a
normal star of the same mass (as done in the simulations of
\citealp{article:hurley_m67}) produces an evolution track that is not
bright enough. Both approximations overestimate the lifetime of the
collision product. These considerations will affect the predicted
colour-magnitude diagram distribution of collisional blue stragglers
from a cluster simulation.

Our code is suitable for a systematic exploration of the wide
parameter space of collisions in clusters of different ages. This will
be the topic of future papers. Eventually we hope to integrate our
code into a full $N$-body code to allow for more realistic and
self-consistent star cluster simulations.


\begin{acknowledgements}
We thank the referee, Alison Sills, for useful comments that improved this
paper.
EG acknowledges support from NWO under grant 614.000.303
\end{acknowledgements}

\begin{appendix}

\section{Modifications to the evolution code}
It is a feature of the STARS code that it solves the stellar structure
equations simultaneously with the reaction-diffusion equations for the
different abundances on a moving mesh. The code is normally fast and reliable,
but we have found that it sometimes has difficulty evolving our merger
remnants and have come up with a scheme that helps it evolve through
``difficult'' timesteps.

Consider the set of independent variables $\vec H$ that represents a solution
to the stellar structure equations at time $t$. The problem is then to find
the changes $\Delta \vec H$ such that $\vec H + \Delta \vec H$  represents
the solution at time $t + \Delta t$. An initial guess for $\Delta H$ can be
taken from the previous timestep and then improved by iteration in a Henyey-like
solver. If no solution can be found, a smaller timestep can be tried.
In some cases, this leads to a runaway situation where repeated convergence
failures cause the timestep to drop until it drops below a threshold value
and the code aborts.

The cause of the convergence failure is that the initial guess for the
corrections $\Delta \vec H$ is not close enough to the desired corrections.
We have looked for ways to improve the initial guess in case of convergence
failure.

Often it is possible to identify the terms in the equations that cause 
difficulty. A common example are the diffusion terms in the reaction-diffusion
equations,
\beq
\diff{X_i}{}{t} = -\frac{1}{\rho r^2} \nabla \rho r^2 \sigma \nabla X_i + R_i,
\eeq
where $X_i$ represents the abundance of species $i$, $\sigma$ is the sum
of all diffusion coefficients affecting the composition and $R_i$ is the
production (or destruction) rate of species $i$ due to nuclear reactions.
In case of convergence problems, it can help te eliminate or reduce (``relax'')
the diffusion coefficient $\sigma$. The resulting corrections $\Delta \vec H'$
are not the final corrections, but they might be a better first guess than
the values used previously.

Convective mixing is the most common example where our above relaxation scheme
is useful, but it is by no means the only one. Other examples where we have
found it useful in our code are the nuclear energy generation rate (which is 
then relaxed from the value at the previous timestep), the mass loss rate and
advection terms in the luminosity equation.

\end{appendix}

\bibliographystyle{aa}
\bibliography{sm67bss}

\begin{thebibliography}{43}
\expandafter\ifx\csname natexlab\endcsname\relax\def\natexlab#1{#1}\fi

\bibitem[{{Aarseth}(1999)}]{1999PASP..111.1333A}
{Aarseth}, S.~J. 1999, \pasp, 111, 1333

\bibitem[{{Ahumada} \& {Lapasset}(1995)}]{1995A&AS..109..375A}
{Ahumada}, J. \& {Lapasset}, E. 1995, \aaps, 109, 375

\bibitem[{{Alexander} \&
  {Ferguson}(1994)}]{article:alexander_ferguson_lowtopac}
{Alexander}, D.~R. \& {Ferguson}, J.~W. 1994, \apj, 437, 879

\bibitem[{{Anders} \& {Grevesse}(1989)}]{article:anders_and_grevesse_abund}
{Anders}, E. \& {Grevesse}, N. 1989, \gca, 53, 197

\bibitem[{{Bailyn} \& {Pinsonneault}(1995)}]{article:bailyn_pinsonneault}
{Bailyn}, C.~D. \& {Pinsonneault}, M.~H. 1995, \apj, 439, 705

\bibitem[{{B\"ohm-Vitense}(1958)}]{article:bohm-vitense_convection}
{B\"ohm-Vitense}, E. 1958, ZsAp, 46, 108

\bibitem[{{Caughlan} \&
  {Fowler}(1988)}]{article:caughlan_and_fowler_reactionrates}
{Caughlan}, G.~R. \& {Fowler}, W.~A. 1988, Atomic Data and Nuclear Data Tables,
  40, 283

\bibitem[{{Caughlan} {et~al.}(1985){Caughlan}, {Fowler}, {Harris}, \&
  {Zimmerman}}]{article:caughlan_reactionrates}
{Caughlan}, G.~R., {Fowler}, W.~A., {Harris}, M.~J., \& {Zimmerman}, B.~A.
  1985, Atomic Data and Nuclear Data Tables, 32, 197

\bibitem[{{Eggleton}(1971)}]{article:eggleton_evlowmass}
{Eggleton}, P.~P. 1971, \mnras, 151, 351

\bibitem[{{Eggleton}(1972)}]{article:eggleton_mixingproc}
{Eggleton}, P.~P. 1972, \mnras, 156, 361

\bibitem[{{Glebbeek} \& {Pols}(2008)}]{GlebbeekPols2008}
{Glebbeek}, E. \& {Pols}, O.~R. 2008, A\&A submitted

\bibitem[{{Gratton} {et~al.}(2000){Gratton}, {Sneden}, {Carretta}, \&
  {Bragaglia}}]{2000A&A...354..169G}
{Gratton}, R.~G., {Sneden}, C., {Carretta}, E., \& {Bragaglia}, A. 2000, \aap,
  354, 169

\bibitem[{{Hills} \& {Day}(1976)}]{1976ApL....17...87H}
{Hills}, J.~G. \& {Day}, C.~A. 1976, \aplett, 17, 87

\bibitem[{{Hurley} {et~al.}(2005){Hurley}, {Pols}, {Aarseth}, \&
  {Tout}}]{article:hurley_m67}
{Hurley}, J.~R., {Pols}, O.~R., {Aarseth}, S.~J., \& {Tout}, C.~A. 2005,
  \mnras, 363, 293

\bibitem[{{Hurley} {et~al.}(2000){Hurley}, {Pols}, \&
  {Tout}}]{article:hurley_sse}
{Hurley}, J.~R., {Pols}, O.~R., \& {Tout}, C.~A. 2000, \mnras, 315, 543

\bibitem[{{Hurley} {et~al.}(2001){Hurley}, {Tout}, {Aarseth}, \&
  {Pols}}]{article:hurley_first_m67}
{Hurley}, J.~R., {Tout}, C.~A., {Aarseth}, S.~J., \& {Pols}, O.~R. 2001,
  \mnras, 323, 630

\bibitem[{{Hurley} {et~al.}(2002){Hurley}, {Tout}, \&
  {Pols}}]{article:hurley_bse}
{Hurley}, J.~R., {Tout}, C.~A., \& {Pols}, O.~R. 2002, \mnras, 329, 897

\bibitem[{{Johnson} \& {Sandage}(1955)}]{1955ApJ...121..616J}
{Johnson}, H.~L. \& {Sandage}, A.~R. 1955, \apj, 121, 616

\bibitem[{{Kippenhahn} {et~al.}(1980){Kippenhahn}, {Ruschenplatt}, \&
  {Thomas}}]{article:kippenhahn_thermohalinemixing}
{Kippenhahn}, R., {Ruschenplatt}, G., \& {Thomas}, H.-C. 1980, \aap, 91, 175

\bibitem[{{Kippenhahn} \& {Weigert}(1990)}]{book:kippenhahn_weigert}
{Kippenhahn}, R. \& {Weigert}, A. 1990, {Stellar Structure and Evolution}
  (Stellar Structure and Evolution, XVI, 468 pp.~192 figs..~ Springer-Verlag
  Berlin Heidelberg New York.~Also Astronomy and Astrophysics Library)

\bibitem[{{Kurucz}(1992)}]{article:kurucz}
{Kurucz}, R.~L. 1992, in IAU Symposium, Vol. 149, The Stellar Populations of
  Galaxies, ed. B.~{Barbuy} \& A.~{Renzini}, 225--+

\bibitem[{{Landau} \& {Lifshitz}(1980)}]{book:landau_ligshitz_statphys1}
{Landau}, L.~D. \& {Lifshitz}, E.~M. 1980, {Statistical physics. Pt.1, Pt.2}
  (Course of theoretical physics, Pergamon International Library of Science,
  Technology, Engineering and Social Studies, Oxford: Pergamon Press,
  1980|c1980, 3rd rev.and enlarg.~ed.)

\bibitem[{{Lejeune} {et~al.}(1997){Lejeune}, {Cuisinier}, \&
  {Buser}}]{article:lejeune_calibration_of_spectra}
{Lejeune}, T., {Cuisinier}, F., \& {Buser}, R. 1997, \aaps, 125, 229

\bibitem[{{Lombardi} {et~al.}(1995){Lombardi}, {Rasio}, \&
  {Shapiro}}]{1995ApJ...445L.117L}
{Lombardi}, Jr., J.~C., {Rasio}, F.~A., \& {Shapiro}, S.~L. 1995, \apjl, 445,
  L117

\bibitem[{{Lombardi} {et~al.}(1996){Lombardi}, {Rasio}, \&
  {Shapiro}}]{1996ApJ...468..797L}
{Lombardi}, Jr., J.~C., {Rasio}, F.~A., \& {Shapiro}, S.~L. 1996, \apj, 468,
  797

\bibitem[{{Lombardi} {et~al.}(2002){Lombardi}, {Warren}, {Rasio}, {Sills}, \&
  {Warren}}]{article:lombardi_mmas}
{Lombardi}, Jr., J.~C., {Warren}, J.~S., {Rasio}, F.~A., {Sills}, A., \&
  {Warren}, A.~R. 2002, \apj, 568, 939

\bibitem[{{Mathys}(1991)}]{mathys_bss_m67}
{Mathys}, G. 1991, \aap, 245, 467

\bibitem[{{Peterson} {et~al.}(1984){Peterson}, {Carney}, \&
  {Latham}}]{1984ApJ...279..237P}
{Peterson}, R.~C., {Carney}, B.~W., \& {Latham}, D.~W. 1984, \apj, 279, 237

\bibitem[{{Pols} {et~al.}(1998){Pols}, {Schroder}, {Hurley}, {Tout}, \&
  {Eggleton}}]{article:pols_evmodels}
{Pols}, O.~R., {Schroder}, K.-P., {Hurley}, J.~R., {Tout}, C.~A., \&
  {Eggleton}, P.~P. 1998, \mnras, 298, 525

\bibitem[{{Pols} {et~al.}(1995){Pols}, {Tout}, {Eggleton}, \&
  {Han}}]{article:pols_approxphys}
{Pols}, O.~R., {Tout}, C.~A., {Eggleton}, P.~P., \& {Han}, Z. 1995, \mnras,
  274, 964

\bibitem[{{Rogers} \& {Iglesias}(1992)}]{article:opal1992}
{Rogers}, F.~J. \& {Iglesias}, C.~A. 1992, \apjs, 79, 507

\bibitem[{{Sandage}(1953)}]{sandage_m3}
{Sandage}, A.~R. 1953, \aj, 58, 61

\bibitem[{{Sandquist} {et~al.}(1997){Sandquist}, {Bolte}, \&
  {Hernquist}}]{article:sandquist_bolte_hernquist}
{Sandquist}, E.~L., {Bolte}, M., \& {Hernquist}, L. 1997, \apj, 477, 335

\bibitem[{{Schr\"oder} {et~al.}(1997){Schr\"oder}, {Pols}, \&
  {Eggleton}}]{article:schroeder_overshooting}
{Schr\"oder}, K.-P., {Pols}, O.~R., \& {Eggleton}, P.~P. 1997, \mnras, 285, 696

\bibitem[{{Sills} {et~al.}(2005){Sills}, {Adams}, \&
  {Davies}}]{sills_evcolprod3}
{Sills}, A., {Adams}, T., \& {Davies}, M.~B. 2005, \mnras, 358, 716

\bibitem[{{Sills} {et~al.}(2002){Sills}, {Adams}, {Davies}, \&
  {Bate}}]{2002MNRAS.332...49S}
{Sills}, A., {Adams}, T., {Davies}, M.~B., \& {Bate}, M.~R. 2002, \mnras, 332,
  49

\bibitem[{{Sills} {et~al.}(1995){Sills}, {Bailyn}, \&
  {Demarque}}]{1995ApJ...455L.163P}
{Sills}, A., {Bailyn}, C.~D., \& {Demarque}, P. 1995, \apjl, 455, L163+

\bibitem[{{Sills} {et~al.}(2001){Sills}, {Faber}, {Lombardi}, {Rasio}, \&
  {Warren}}]{sills_evcolprod2}
{Sills}, A., {Faber}, J.~A., {Lombardi}, Jr., J.~C., {Rasio}, F.~A., \&
  {Warren}, A.~R. 2001, \apj, 548, 323

\bibitem[{{Sills} {et~al.}(1997){Sills}, {Lombardi}, {Bailyn}, {Demarque},
  {Rasio}, \& {Shapiro}}]{article:sills_evcolprod1}
{Sills}, A., {Lombardi}, Jr., J.~C., {Bailyn}, C.~D., {et~al.} 1997, \apj, 487,
  290

\bibitem[{{Stancliffe} {et~al.}(2007){Stancliffe}, {Glebbeek}, {Izzard}, \&
  {Pols}}]{article:stancliffe_thmixing}
{Stancliffe}, R.~J., {Glebbeek}, E., {Izzard}, R.~G., \& {Pols}, O.~R. 2007,
  \aap, 464, L57

\bibitem[{{Stryker}(1993)}]{1993PASP..105.1081S}
{Stryker}, L.~L. 1993, \pasp, 105, 1081

\bibitem[{{Tout} {et~al.}(1997){Tout}, {Aarseth}, {Pols}, \&
  {Eggleton}}]{article:tout_evolution_models}
{Tout}, C.~A., {Aarseth}, S.~J., {Pols}, O.~R., \& {Eggleton}, P.~P. 1997,
  \mnras, 291, 732

\bibitem[{{Ulrich}(1972)}]{article:ulrich_thermohaline}
{Ulrich}, R.~K. 1972, \apj, 172, 165

\end{thebibliography}

\end{document}